\documentstyle[12pt, fleqn]{article}
\begin{document}

\title{\bf  Nuclear Medium Effects in the Relativistic Treatment of
            Quasifree Electron Scattering\thanks{
            Work supported in part by the Natural Sciences and
            Engineering Research Council of Canada} }

\author{\bf  M. Hedayati-Poor, J.I. Johansson and H.S. Sherif  \\
Department of Physics, University of Alberta \\
Edmonton,  Alberta, Canada T6G 2J1}

\date{\today}

\maketitle

\begin{abstract}

Non-relativistic reduction of the S-matrix for the quasifree
electron scattering process $A\left(~e, e'p~\right)A-1$ is studied
in order to understand the source of differences between
non-relativistic and relativistic models.
We perform an effective Pauli reduction on the relativistic
expression for the S-matrix in the one-photon exchange
approximation.
The reduction is applied to the nucleon current only; the
electrons are treated fully relativistically.
An expansion of the amplitude results in a power series in the
nuclear potentials.
The series is found to converge rapidly only if the nuclear
potentials are included in the nuclear current operator.
The results can be cast in a form which reproduces the
non-relativistic amplitudes in the limit that the potentials are
removed from the nuclear current operator.
Large differences can be found between calculations which do and do
not include the nuclear potentials in the different orders of the
nuclear current operator.
In the high missing momentum region we find that the
non-relativistic
calculations with potentials included in the nuclear current up to
second order give results which are close to those of the fully
relativistic calculation.
This behavior is an indication of the importance of the medium
modifications of the nuclear currents in this model, which are
naturally built into the relativistic treatment of the reaction.

\end{abstract}


\newpage

\section{Introduction}      \label{intro}

Electromagnetic probes provide invaluable information about nuclear
structure.
The  quasifree process $(~e, e'p~)$ has been used extensively to
study proton hole states and to determine single particle
spectroscopic factors \cite{SJ}.
This reaction is advantageous because the electromagnetic
interaction is known; the relative weakness of the reaction
permits the probe to interact almost uniformly through the entire
nucleus,
and the first order of perturbation theory should provide
an adequate description of the process.
Coincidence measurements of the $(~e, e'p~)$ reaction can provide
detailed information about the single particle structure of the
nucleus over a wide range of momentum transfer.

The $(~e, e'p~)$ reaction has been widely studied both
non-relativistically \cite{BGP,ML} and relativistically
\cite{J,JO,JU}, and there are some discrepancies between the results
of these investigations.
Both analyses begin with a lagrangian which allows for the
interaction
of the photon with both electrons and protons.
Non-relativistic analyses involve the reduction of the free
electron-proton interaction to a form involving two-component
spinors for the nucleon.
This results in a hamiltonian which is expanded in powers of $1/M$
where $M$ is the nucleon  mass \cite{BGP}.
The resulting interaction hamiltonian is sandwiched between
Schr\"{o}dinger wave functions describing the nucleons in order to
form the nuclear current.
Relativistic analyses are based on the Feynman diagram for
one-photon exchange between the projectile electron and a proton which
is imbedded in the nucleus.
The electrons and nucleons are all described relativistically as spin
1/2 objects via the Dirac equation containing  appropriate potentials
\cite{J,JO,JU}.
A long-standing problem in quasifree electron scattering has been
that the spectroscopic factors extracted from non-relativistic
analyses are smaller than expected from shell model calculations.
Spectroscopic factors which are found on the basis of the
relativistic approach are generally larger than those found via the
non-relativistic approach \cite{J,JU}.

Several groups have attempted to understand the underlying
differences between these two approaches.
This mainly involved looking at the sensitivity of quasifree electron
scattering calculations to different optical potentials and
renormalizations of the continuum wave function
\cite{BGPC87,Ud93,JO94}.
This concentration on optical potentials was largely a result of the
improvement in the description of proton elastic scattering
observables
via Dirac phenomenology over the standard non-relativistic optical
model description.
Boffi et al. \cite{BGPC87} have multiplied the non-relativistic
continuum wave function by a potential-dependent factor
$\left\{ 1 + \left[ S\left( r \right) - V\left( r \right) \right] /
\left( E + M \right) \right\}^{1/2}$,
where $S\left( r \right)$ is the Dirac scalar potential and
$V\left( r \right)$ is the vector potential.
This modification essentially changes the two-component
Schr\"{o}dinger wave function into the upper component of the Dirac
wave function, while no other change is made in the non-relativistic
calculation.
They find that extracted occupation probabilities are larger than
those obtained from the unmodified  non-relativistic analysis.
The analysis of Udias et al. \cite{Ud93} replaces the
non-relativistic
bound state wave function with the upper component of a Dirac wave
function, and the non-relativistic continuum wave function is
modified
by factors of the same shape as the factor used by Boffi et al.
The continuum wave function in this case is generated from
Schr\"{o}dinger-equivalent potentials \cite{SSC86}.
The nuclear current operators are obtained in the standard way by
expansion to order $1/M^{4}$.
Their "non-relativistic" calculations then involve non-relativistic
nuclear current operators surrounded by the upper components of
Dirac wave functions.
With these choices little difference is found between the
relativistic and "non-relativistic" calculations.
Their conclusion is that differences in observed cross sections are
not due to  non-relativistic reduction, but to the choice of optical
potential.
Jin and Onley \cite{JO94} have presented a model which can take
either
relativistic or non-relativistic optical potentials while keeping
other aspects of the calculation the same.
They find that different optical potentials can change the results
by as much as 14\%.

These results clearly demonstrate the variability due to final state
interactions, however, the issue is clouded by the occasional use of
upper components of Dirac wave functions in a non-relativistic
calculation.
We believe that the essential difference between relativistic and
non-relativistic approaches are not just in the changes in the
optical
potentials; these are usually phenomenological and equivalent
potentials can always be found.
Rather the essential difference is in the appearance of the nuclear
potentials in the nuclear current operators when the relativistic
amplitude is reduced to a non-relativistic form.
Such medium effects on the nuclear currents are absent in the
standard non-relativistic calculations.

In this paper we study the differences between the relativistic
and non-relativistic approaches in calculating the amplitude for the
$(~e, e'p~)$ reaction.
We do this through an effective Pauli reduction of the relativistic
transition amplitude \cite{HH}.
An expansion of the amplitude in powers of
$\left( E + M \right)^{-1}$ allows us to recover a non-relativistic
limit, which matches the standard non-relativistic calculations,
with the difference that optical potentials used to generate the
distorted waves are exactly equivalent to those used in the
relativistic calculations.
The main difference, as mentioned above is that the nuclear currents
are potential dependent.
We compare the two approaches and thus explain why they can still
give different values for the extracted spectroscopic factors, even
when equivalent optical potentials are used.

We introduce the relativistic amplitude for quasifree electron
scattering in  section \ref{rel-amp}.
Section  \ref{Pauli} outlines the Pauli reduction of the
amplitude and some of its relevant features are discussed in section
\ref{E-results}.
The non-relativistic limit is discussed in section \ref{nonrel}.
In section \ref{n-results} we compare our non-relativistic
calculations with and without nuclear potentials in the nuclear
current operators, to the results of the fully relativistic
calculations.
Our conclusions are given in section \ref{concl}.

\section{Relativistic Amplitude}   \label{rel-amp}

We consider the one photon exchange model for the  $(~e, e'p~)$
process \cite{JU}, in which a photon is exchanged between the
incident electron and a target proton.
The struck proton is detected in coincidence with the final state
electron.
In this paper we are interested in the differences between the
relativistic and non-relativistic treatment of the hadronic part
the $(~e, e'p~)$ reaction.
In the course of this discussion we do not include the Coulomb
interaction in the leptonic part of the S-matrix since this will
only be important for heavy nuclei \cite{J,GP87}.
We will therefore not discuss any nuclei heavier than $^{90}Zr$ in
this work.

The relativistic expression for the S-matrix describing the
quasifree electron scattering process $(~e, e'p~)$ in the
distorted wave Born approximation (DWBA) is \cite{JU}
\begin{eqnarray}
S_{fi} &=& \frac{- ie^2 }{(2\pi)^{17/2}}
           \left[ \frac{M \; m_e^{2}}{E_c E_f E_i} \right ]^{1/2}
      \sum_{J_B M_B} { \left(J_f, J_B;M_f, M_B| J_i, M_i \right)}
         \nonumber \\
       &  &\times{ \left[ {\cal S}_{J_i J_f} (J_B) \right] }^{1/2}
                   \int d^4x \, d^4y \, d^4q \,
                   J_{e\mu}(y) \frac{e^{-iq\cdot( x-y)}}
                                    {q^2 + i\epsilon} J^\mu_N(x) ,
         \label{f1}
\end{eqnarray}
where ${\cal S}_{J_i J_f} (J_B)$ is the spectroscopic factor and
$J_{e}^{\mu}$ and $J^\mu_N$ are electron and nuclear currents
respectively. $M$ is the nucleon mass and  $E_c$  is the energy of
the outgoing proton.
The electron current is given by
\begin{eqnarray}
J^\mu_e\left( y \right)
        = \overline{\psi}_{e_f }\left( y \right) \gamma^\mu
                   \psi_{e_i }\left( y \right) ,
          \label{f2}
\end{eqnarray}
where $\psi_{e_i }$ and $\psi_{e_f }$ are the initial and final
Dirac spinors for electrons.
The electron wave functions are taken to be free Dirac spinors and
the integration at the electron vertex can then be done analytically.
This also allows the momentum integration for the photon propagator
to be done, leaving one four-dimensional integration at the
nucleon vertex.
The nuclear current is similarly given by
\begin{eqnarray}
J^\mu_N \left( x \right)
      = \overline{\psi}_{N_f}\left( x \right) j^\mu_N
          \psi_{N_b}\left( x \right) ,
          \label{f3}
\end{eqnarray}
where the nuclear current operator $j^\mu_N$ is the choice $cc2$
discussed by de Forest \cite{TDFJ}
\begin{eqnarray}
j^\mu_N = F_1\left( q^2 \right) \gamma^\mu
           +  \frac{ i \kappa F_2\left( q^2 \right)} {2M}
              \sigma^{\mu\nu}q_{\nu} .
           \label{f4}
\end{eqnarray}
$F_1\left( q^2 \right)$ and $F_2\left( q^2 \right)$ are nucleon
form factors and are functions of the four-momentum squared of the
exchanged photon, which couples to the nucleons.
We have $q^{\mu}= k^{\mu}_{i} - k^{\mu}_{f}$ where $k^{\mu}_{i}$
and $k^{\mu}_{f}$  are the momenta of the initial and final
electrons respectively.
The matrix $\sigma^{\mu\nu}$ is formed from the Dirac
$\gamma$-matrices in the standard way as \cite{BD}
\begin{eqnarray}
\sigma^{\mu\nu}
     = \frac{i}{2} \left(   \gamma^\mu \gamma^\nu
                          - \gamma^\nu \gamma^\mu \right) .
            \label{f5}
\end{eqnarray}
The integration over $d^4q$ in Eq. (\ref{f1}) is associated with
the propagator of the exchanged photon.

The electrons are described by positive energy Dirac spinors, and
the integration over coordinates at the electron vertex, in the
S-matrix of equation in Eq. (\ref{f1}), can be done analytically to
yield a Dirac $\delta$-function giving energy and momentum
conservation at the vertex so $q = k_{i}^{e} - k_{f}^{e}$.
The resulting $\delta$-function fixes the momentum of the
intermediate photon so the integration over that momentum
is done trivially. The integration over the time coordinate at the
nucleon vertex can then be done to yield a $\delta$-function
providing overall energy conservation.
The S-matrix can then be cast in the form
\begin{eqnarray}
S_{fi} &=& \frac{ - i e^2 } { \left( 2\pi \right)^{7/2} }
            \frac{1} { q_\gamma^{2} }
            \left[ \frac{M \; m_e^{2}}{E_c E_f E_i} \right ]^{1/2}
            \delta \left( E_c + E_f - E_b - E_i \right)
         \nonumber \\
      & & \times \sum_{ J_B M_B}{
                 \left( J_f, J_B; M_f, M_B| J_i, M_i \right)
              { \left[ {\cal S}_{J_i J_f} (J_B) \right] }^{1/2}
              Z_{\nu_{f} \nu_{i}}^{\mu M_B} } ,
\end{eqnarray}
where $Z_{\nu_{f} \nu_{i}}^{\mu M_B}$ is a function of the initial
and final spin projections, momenta, etc.
Specifically we have at this point
\begin{eqnarray}
  Z_{\nu_{f} \nu_{i}}^{\mu M_B} = e_{\nu_{f} \nu_{i}}^{\alpha}
        \int d^3x \;
        \Psi_{\mu}^\dagger \left( k_{p}, \mbox{\boldmath{$x$}} \right)
        \Gamma_{\alpha}
        \Psi_{J_{B}, M_{B}} \left( \mbox{\boldmath{$x$}} \right)
        \exp \left( i \mbox{\boldmath{$q$}} \cdot
                      \mbox{\boldmath{$x$}} \right)     .
\end{eqnarray}
where the $4 \times 4$ matrix operating on the nucleon spinors is
\begin{eqnarray}
   \Gamma_{\alpha}
               = \gamma_{0} \left[ F_1\left( q^2 \right)
                                   \gamma_{\alpha}
                                   + \frac{i \kappa F_2( q^2)}
                                     {2M} \sigma_{\alpha \nu}
                                     q^\nu \right]   .
\end{eqnarray}
The four-vector which comes from the electron vertex is
\begin{eqnarray}
   e_{\nu_{f} \nu_{i}}^{\alpha}
       &=& \left[ \frac{E_{f} + m_{e}} {2m_{e}}
                  \frac{E_{i} + m_{e}} {2m_{e}} \right]^{1/2}
         \nonumber \\
       & &  \times \left< 1/2, \nu_{f} \right|
            \left[ 1, \matrix{ \mbox{\boldmath{$\sigma$}} \cdot
                      \mbox{\boldmath{$k$}}_{f} \cr
                      \overline{E_{f} + m_{e}} } \right]
            \gamma_{0} \gamma^{\alpha}
            \left[ \matrix{ 1 \cr
                            \mbox{\boldmath{$\sigma$}} \cdot
                            \mbox{\boldmath{$k$}}_{i} \cr
                            \overline{E_{i} + m_{e}} } \right]
            \left| 1/2, \nu_{i} \right>         .
\end{eqnarray}
and this depends on the energies and momenta of the initial and
final electrons as well as their spin projections.
Useful details of the Fock space calculations and the expansion of
the Dirac wave function in partial waves can be found in the paper
by Johansson and Sherif \cite{JS94}.
When the appropriate factors of $\hbar$ and $c$ are included,
the relativistic expression for the triple differential cross
section is related to $Z_{\nu_{f} \nu_{i}}^{\mu M_B}$ by
\begin{eqnarray}
\frac{d^3\sigma}{d\Omega_{e} d\Omega_p dE_p}  &=&
  \frac{2} {\left( 2 \pi \right)^{3}}  \frac{\alpha^{2}} {\hbar c}
   { \left[  \frac{ \left( m_e c^{2} \right)^{2} M c^{2} \; p_p c}
   { \left( q_\gamma c \right)^{4} } \frac{ p_f c} { p_i c}\right] }
  \nonumber
   \\  & & \times \sum_{ J_B M_B \mu  \nu_{f} \nu_{i} }
    { \frac{ {\cal S}_{J_i J_f} (J_B) }{ 2J_B + 1 } }
    { | Z_{\nu_{f} \nu_{i}}^{\mu M_B} | }^2 ,
  \label{cross}
\end{eqnarray}
where $\nu_i$ and $\nu_f$ are the spin projections of the incoming
and outgoing electrons respectively, while $M_B$ and $\mu$ are the
spin projections of the bound and continuum protons.

The cross section for quasifree electron scattering in the plane
wave impulse approximation can be written in a factorized form as
the product of three parts \cite{TDFJ,BGP}: a kinematic factor,
the cross section for the elementary process $e+p\rightarrow e'+p'$,
which is evaluated off-shell,  and finally a function of the
energy and momentum of the nucleon inside the nucleus referred to as
the spectral function.

In the following we will discuss results of calculations of the
spectral function, proton polarization and an asymmetry parameter.
The spectral function is obtained from the cross section given above
by dividing by a kinematic factor and the cross section for the
elementary process for $e+p \rightarrow e+p$.
We write \cite{TDFJ,BGP}:
\begin{eqnarray}
S\left( p_{m} \right) = \frac{ \frac{ d^3\sigma}
    {  d\Omega_{e} d\Omega_c dE_c}  }
    {  E_c p_c \left. \frac{d\sigma}{d\Omega_{e}} \right|_{free}  },
\end{eqnarray}
where $E_c$ and $p_c$ are the energy and momentum of the final state
proton, and $p_m$ is the missing momentum, i.e. the momentum of the
bound nucleon in the initial state.
The free cross section is calculated using the nucleon current
operator
of Eq. (\ref{f4}), and is evaluated using the kinematics of the
quasifree process, i.e. off-shell.
Note that the experimental data are divided by the elementary cross
section $cc1$ of de Forest \cite{TDFJ},
while we use the nuclear current operator which leads to his cross
section $cc2$, throughout this work.
We are not concerned with detailed comparison with experimental
data in this work, so we retain a consistent approach by  using
the same form for the current operator in the calculation of the
quasifree S-matrix and the elementary process.

The polarization of the final state proton is given by
\begin{eqnarray}
P = -2 \; \frac{ \mbox{Im}
      \sum_{ M_B \nu_{f} \nu_{i} } { Z_{\nu_{f} \nu_{i}}^{1/2 M_B}
    { \left[ Z_{\nu_{f} \nu_{i}}^{-1/2 M_B} \right] }^{\ast} }  }
    { \sum_{M_B \mu  \nu_{f} \nu_{i} }
    {    { \left| Z_{\nu_{f} \nu_{i}}^{\mu M_B} \right| }^{2} } }.
\end{eqnarray}
We also define an asymmetry parameter in the missing momentum which
is calculated from the differential cross sections of Eq. (\ref{cross})
as
\begin{eqnarray}
{ \cal A } \left( p_m \right)
        = \frac{   d^{3} \sigma \left( p_m > 0 \right)
                 - d^{3} \sigma \left( p_m < 0 \right)  }
               {   d^{3} \sigma \left( p_m > 0\right)
                 + d^{3} \sigma \left( p_m < 0 \right)  } .
       \label{f15}
\end{eqnarray}
This asymmetry is similar to the parameter defined by Bianconi,
Boffi and Kharzeev \cite{BBK94}.

\section{Pauli Reduction}         \label{Pauli}

We now perform the effective Pauli reduction \cite{HH} on the
hadronic part of  the amplitude only; the electrons are treated
relativistically throughout.
Consider the nuclear current of Eq. (\ref{f3}) above.
Using Eq. (\ref{f4}) one can write:
\begin{eqnarray}
J^\mu \left( x \right)
      = \overline{\psi}_{N_{f}} \left(k_{p}, x \right)
        \left[ F_1\left( q^2 \right) \gamma^\mu
  + \frac{i \kappa F_2( q^2)} {2M} \sigma^{\mu\nu} q_\nu \right]
     \psi_{N_b}\left( x \right) .
        \label{f6}
\end{eqnarray}

The Dirac spinors may be written in terms of their upper
components as
\begin{eqnarray}
\psi_{N} = \left[   \matrix{   1 \cr
                    \mbox{\boldmath{$\sigma$}} \cdot
                    \mbox{\boldmath{$ p$}} \vspace{0 mm} \cr
                    \overline{M + E + S\left( r \right)
                                - V\left( r \right)}   \cr   }
                                                \right] u ,
          \label{f7}
\end{eqnarray}
where $S\left( r \right)$ and $V\left( r \right)$ are the scalar
and vector potentials, respectively, for either the bound or
final state nucleons.
The energy of the nucleon is $E$, and the associated momentum
operator is $\mbox{\boldmath{$ p$}}$.
The upper component of the Dirac spinor $u$ is related to a
Schr\"{o}dinger-like wave function $\Psi_{Sch}$ by \cite{CHM}
\begin{eqnarray}
   u = D^{ \frac{1}{2} } \Psi_{Sch} \quad
      \mbox{where}
      \quad D\left( r \right)
        = \frac{E + M + S\left( r \right) - V\left( r \right)}
                  {E + M}.
     \label{f8}
\end{eqnarray}
Note that the two-component wave function $\Psi_{Sch}$ is a solution
of the Schr\"{o}\-dinger equation used in the ordinary
non-relativistic calculations, i.e. containing central and spin-orbit
potentials, but these potentials are "equivalent potentials" meaning
that they are functions of  the vector  and scalar potentials of the
original Dirac equation, as well as containing explicit energy
dependence.
For the continuum nucleon the Dirac potentials result in an improved
description of nucleon-nucleus elastic scattering data \cite{COPE},
while for the bound state the Dirac potentials offer a slightly better
description of the spin-orbit splitting  than those used in earlier
non-relativistic calculations \cite{HS}.

The relativistic nuclear current of Eq. (\ref{f6}) can, with the help
of Eq. (\ref{f8}), be written in the form
\begin{eqnarray}
J^\mu &=& \Psi_{Sch,c}^{\dagger} \left\{ \matrix{ \cr \cr \cr }
                         D_c^{1/2}\left( r \right) \left[  1, \;
                        \frac{ \mbox{\boldmath{$\sigma$}} \cdot
                               \mbox{\boldmath{$p$}} }
                               {M + E_c + S_c\left( r \right)
                                - V_c\left( r \right)}
                                            \right] \right.
                    \nonumber \\
   & & \hspace{18.5 mm} \times \;\; \gamma^o \left[ F_1 \gamma^\mu
         + F_2 \frac{i \kappa }{2M} \sigma^{\mu\nu}q_\nu \right]
                  \label{f9} \\
  & & \hspace{18.5 mm} \times \;\;\;\;\;
            \left. \left[   \matrix{   1   \cr
            \mbox{\boldmath{$\sigma$}} \cdot
            \mbox{\boldmath{$ p$}}  \vspace{0 mm}    \cr
            \overline{M + E_b + S_b\left( r \right)
                       - V_b\left( r \right)}    \cr   } \right]
             D_b^{1/2}\left( r \right) \right\} \Psi_{Sch,b} .
         \nonumber
\end{eqnarray}
We now perform an expansion of the object between the braces
of Eq. (\ref{f9}). The usual representation of the Dirac
$\gamma$-matrices is used \cite{BD} to write the
$4 \times 4$ operator in terms of $2 \times 2$ Pauli matrices.
The matrix multiplications are performed and a $2 \times 2$
operator results. The radial function
$D^{1/2}\left( r \right)$  from Eq. (\ref{f8}) and the
factor $\left[ E + M + S\left( r \right) - V\left( r \right)
\right]^{-1}$ coming from both the bound and continuum
wave functions are then expanded in powers of
$\left( E + M \right)^{-1}$.
This procedure leads to a sum of reduced nuclear current operators
for each of the contributing orders:
\begin{eqnarray}
J^\mu \left( x \right) = \Psi_{Sch, c}^\dagger\left( x \right)
           \left[   j^{\mu ^ {(0)}} +  j^{\mu ^ {(1)}}
                  + j^{\mu ^ {(2)}} + \cdots        \right]
           \Psi_{Sch, b}\left( x \right) .
          \label{f10}
\end{eqnarray}
The reduced current operators can be written in terms of time-like
and space-like components as
\begin{eqnarray}
j^{0 ^ {(0)}} = && e F_1 ,
                      \nonumber \\
j^{0 ^ {(1)}} = && \frac{e F_1}{2} \left[ Q_c + Q_b \right] ,
                      \nonumber \\
j^{0 ^ {(2)}} = && eF_1 \left[ \frac{Q_c Q_b}{4}
                   - \frac{Q_c^2 + Q_b^2}{8}
                   + \frac{ \mbox{\boldmath{$\sigma$}} \cdot
                            \mbox{\boldmath{$p$}}}
                          {M + E_c}
                     \frac{\mbox{\boldmath{$\sigma$}} \cdot
                           \mbox{\boldmath{$p$}}}
                          {M + E_b}        \right]
                      \nonumber \\
           &+& \frac{\kappa F_2}{2M}
                   \left[ \frac{ \mbox{\boldmath{$\sigma$}} \cdot
                                 \mbox{\boldmath{$q$}} \;\;
                                 \mbox{\boldmath{$\sigma$}} \cdot
                                 \mbox{\boldmath{$p$}}}
                               {M + E_b}
                          - \frac{ \mbox{\boldmath{$\sigma$}} \cdot
                                   \mbox{\boldmath{$p$}} \;\;
                                   \mbox{\boldmath{$\sigma$}} \cdot
                                   \mbox{\boldmath{$q$}}}
                                 {M + E_c}             \right] ,
                       \nonumber \\
{\bf j}^{(0)} = && 0 ,
                      \nonumber \\
{\bf j}^{(1)} = & &\frac{ie\kappa F_2}  {2M} \;
                            \mbox{\boldmath{$\sigma$}} \times
                            \mbox{\boldmath{$q$}}
               + eF_1 \left[ \frac{ \mbox{\boldmath{$\sigma$}} \;\;
                                    \mbox{\boldmath{$\sigma$}} \cdot
                                    \mbox{\boldmath{$p$}}}
                                  {M+E_b}
               + \frac{ \mbox{\boldmath{$\sigma$}} \cdot
                        \mbox{\boldmath{$p$}} \;\;
                        \mbox{\boldmath{$\sigma$}}}
                      {M + E_c}                         \right] ,
                          \nonumber \\
{\bf j}^{(2)} = &&\frac{ ie \kappa F_2}{4M} \left[Q_b + Q_c \right]
                            \mbox{\boldmath{$\sigma$}} \times
                            \mbox{\boldmath{$q$}}
                + \frac{ e F_1 Q_c } {2}
                   \left[  \frac{ \mbox{\boldmath{$\sigma$}} \;\;
                                  \mbox{\boldmath{$\sigma$}} \cdot
                                  \mbox{\boldmath{$p$}}}
                                {M+E_b}
                +\frac{ \mbox{\boldmath{$\sigma$}} \cdot
                        \mbox{\boldmath{$p$}} \;\;
                        \mbox{\boldmath{$\sigma$}}}
                      {M+E_c}                           \right]
                          \nonumber \\
              &+& \frac{ e F_1 } {2}
                  \left[ \frac{ \mbox{\boldmath{$\sigma$}}  \;\;
                                \mbox{\boldmath{$\sigma$}} \cdot
                                \mbox{\boldmath{$p$}}}
                              { M + E_b }
                         + \frac{ \mbox{\boldmath{$\sigma$}} \cdot
                                  \mbox{\boldmath{$p$}} \;\;
                                  \mbox{\boldmath{$\sigma$}}}
                                { M + E_c }           \right] Q_b
                           \nonumber \\
              &-& e F_1 \left[ Q_b
                    \frac{\mbox{ \boldmath{$\sigma$}} \;\;
                          \mbox{\boldmath{$\sigma$}} \cdot
                          \mbox{\boldmath{$p$}}}
                         { M + E_b }
                  +\frac{ \mbox{\boldmath{$\sigma$}} \cdot
                          \mbox{\boldmath{$p$}} \;\;
                          \mbox{\boldmath{$\sigma$}}}
                        { M + E_c }               Q_c   \right]
                           \nonumber \\
              &+& \frac{ \kappa F_2 q_o } {2M}
                  \left[ \frac{ \mbox{\boldmath{$\sigma$}} \;\;
                                \mbox{\boldmath{$\sigma$}} \cdot
                                \mbox{\boldmath{$p$}}}
                              { M + E_b }
                    - \frac{ \mbox{\boldmath{$\sigma$}} \cdot
                             \mbox{\boldmath{$p$}} \;\;
                             \mbox{\boldmath{$\sigma$}}}
                           { M + E_c }                 \right]  ,
          \label{f11}
\end{eqnarray}
where we have defined
\begin{eqnarray}
Q\left( r \right)
       = \frac{S\left( r \right) -V\left( r \right) } {E + M} ,
                               \label{f12}
\end{eqnarray}
and the labels $b$ and $c$ refer to the bound and continuum states,
respectively.
Using the current operators from Eq. (\ref{f11}) up to first order
in $\left( E + M \right)^{-1}$, in the S-matrix (\ref{f1}),
we find that for the $(~e, e'p~)$ reaction the S-matrix to first
order in $\left( E + M \right)^{-1}$ reduces to
\begin{eqnarray}
S_{fi}^{\left( 1 \right)}
         &=& \frac{- i e^2 } {(2\pi)^{17/2}}
             \left[ \frac{M \; m_e^{2}}{E_c E_f E_i} \right ]^{1/2}
            \nonumber \\
         & & \times \sum_{J_B M_B} { (J_f, J_B;M_f, M_B| J_i, M_i ) }
             { \left[ {\cal S}_{J_i J_f} (J_B) \right] }^{1/2}
            \nonumber \\
         & & \times \int d^4x \; d^4y \; d^4q \;
                    \frac{e^{-iq \cdot \left( x-y \right)}}
                                { q^2 + i \epsilon }
                    \; \Psi_{Sch, c}^\dagger \left( x \right)
            \nonumber \\
         & & \times \left\{ J_e^0 \left( y \right)
                            F_1\left( q^2 \right)
               \left[ 1 + \frac{1}{2} (Q_c + Q_b) \right] \right.
            \nonumber  \\
         & & \hspace{5 mm}
            - \mbox{\boldmath{$J$}}_e \left( y \right) \cdot
              \left[ i \kappa \mu_N F_2\left( q^2 \right)
              \frac{  \mbox{\boldmath{$\sigma$} }
                      \times \mbox{ \boldmath{$q$} } }
                   {2 M}                                  \right.
                       \nonumber  \\
         & & \left. \left. \hspace{20 mm}
              + F_1\left( q^2 \right)
                \left( \frac{ \mbox{\boldmath{$\sigma$}} \;
                              \mbox{\boldmath{$\sigma$}} \cdot
                              \mbox{\boldmath{$p$}}      }
                            {M + E_b}
              + \frac{ \mbox{\boldmath{$\sigma$}} \cdot
                       \mbox{\boldmath{$p$}}  \;
                       \mbox{\boldmath{$\sigma$}}              }
                     {M + E_c}
         \right) \right] \right\} \Psi_{Sch, b} \left( x \right) .
                       \label{f13}
\end{eqnarray}
Note that we have written this equation in a form in which the
integrations over the electron coordinate and the intermediate
photon momentum have not been done.
The expansion method does not depend on the plane wave approximation
for the electrons, and electron distortions could be included
if desired.
The S-matrix to second order in $\left( E + M \right)^{-1}$ is
similarly found by including the second-order nuclear current as
well.

Note the dependence of the nuclear current operators on the Dirac
vector and scalar potentials (~through the functions $Q_i$~).
This dependence appears in all orders of the reduction scheme.
Thus as we go to a description in terms of the Schr\"{o}dinger-like
wave function for the nucleon, the currents undergo a medium
modification affected via the nuclear potential.
This point is central to the present work. We shall discuss the
traditional non-relativistic limit of the amplitude in section
\ref{nonrel}; but will concentrate in the following section on
clarifying the role of the nuclear potentials in the
convergence properties of the Pauli expansion of the S-matrix.

\section{Convergence of the Expansion}                 \label{E-results}

In this section we discuss the convergence of the expansion obtained
above to the fully relativistic calculation.
In these convergence calculations, all of the factors in the
expansion are the original relativistic ones.
{\em This is not yet equivalent to a standard non-relativistic
calculation}!
The non-relativistic calculations are discussed below.

The calculations of the relativistic S-matrix requires knowledge
of the Dirac wave functions for the bound and continuum states.
For the bound state Hartree bound state wave functions are used
\cite{HS}.
The continuum wave functions for the knocked out proton are obtained
using the energy and A dependent optical potential of Cooper
{\it et al}. \cite{COPE}.
We restrict our discussion to the case of parallel kinematics
\cite{SJ}.
In the diagrams referred to in the following discussion the curves
are labelled
according to their order in $\left( E + M \right)^{-1}$ for the
expansion calculations,
and whether or not the Dirac potentials are included in the nuclear
current operators:
dotted curve -- first order in $\left( E + M \right)^{-1}$ without
Dirac potentials;
dashed curve -- first order with potentials;
dot-dashed curve -- second order without potentials;
dot-dot-dashed curve -- second order with potentials;
solid curve -- fully relativistic calculation.
In doing these comparisons we are attempting to clarify the
convergence of the
expansion and the role of the nuclear potentials (~as they appear
in the
nuclear currents~) in the rate of convergence of this expansion.

Figure 1 shows observables as a function of missing momentum for
the reaction
$^{16}O(~e, e'p~)^{15}N$, 1(a) is the spectral function while 1(b)
is the proton polarization.
The ground state of the residual nucleus, $^{15}N$, is assumed to be
a $1p_{\frac{1}{2}}$ hole.
The energy of the incident electron is 456 MeV, and the kinetic
energy of the
detected proton is fixed at 90 MeV with parallel kinematics.
The relativistic calculations of the spectral function are fitted
to the peak of the data
\cite{L}; the resulting ''spectroscopic factor'' is then used in
all the
other calculations for that particular state.
(~We adopt this simple fitting procedure because our main concern
here is
comparison between the different calculations, rather than a
judicious determination
of the spectroscopic factors.~)
Note that the calculations with potentials included converge
rapidly toward the
fully relativistic results in this case,
with the curve for the second order calculations being very close
the relativistic results over the range of momentum transfers shown.
In calculating the spectral function, the inclusion of the potential
in the first order
interaction terms bring the results closer to the fully relativistic
calculation than the second order without potentials.
It must be stressed that the inclusion of potentials in the
interaction brings the
results close to the fully relativistic results, while the
calculations without potentials are
quite far from the relativistic results and do not show a strong
indication for convergence to the relativistic result.
We have also done similar calculations for the same target but
leaving the residual nucleus in an excited state, as well as using
different targets, namely $^{40}Ca$ and
$^{90}Zr$, with the residual nucleus left in both ground and
excited states.
These calculations show the same behavior as the calculations
shown in Fig. 1.The above results are of course expected on simple mathematical
grounds.
The essential point however, is to shed light on the role of
the appearance of the potentials in the nuclear currents.
We have seen no evidence that expansions that are based on free
vertices (~i.e. no nuclear potentials~),
will converge to the fully relativistic results,
even if calculations are done to higher order in the inverse of
the nucleon mass \cite{BGP}.
This will have implications for the comparisons with the standard
non-relativistic calculations, which we discuss next.

\section{The Non-Relativistic Limit}                \label{nonrel}

The expansion of  the S-matrix in powers of
$\left( E + M \right)^{-1}$ discussed
above does not quite yield the amplitude used in standard
non-relativistic calculations.
Some care must be taken at this point in the discussion to
differentiate between the correct
non-relativistic limit, and the standard operator used in
non-relativistic calculations.
There are three things that must be done in
order to obtain the proper non-relativistic limit from the
relativistic amplitude:
\begin{quote}
i) {\em The bound state wave function must be normalized to unity.}
In the expansion obtained above, the Dirac bound state wave
function is
normalized to unity and the related Schr\"{o}dinger-equivalent
wave function is not.
In the non-relativistic calculations it is the Schr\"{o}dinger-equivalent wave
function that must be normalized. \\
ii) {\em The continuum wave function must be normalized correctly}.
The factors arising from the Dirac field and the normalization of
the Dirac wave
function result in a factor of $\left( E + M \right) / 2E$ being
set equal to one to
obtain the non-relativistic expression for the cross section,
(~this is equivalent to multiplying the right-hand-side of Eq.
(\ref{cross}) by the inverse of this factor~). \\
iii) Finally, to obtain non-relativistic expressions for the
nuclear current operators
from the relativistic expressions of Eq. (\ref{f11}),
{\em the nucleon energies (~both continuum and bound~) are set
equal to the nucleon mass, i.e.  $E \rightarrow M$. }
\end{quote}
It is important to note that
these changes still have not yielded the standard
non-relativistic amplitudes
because the nuclear current operators at this stage contain the
Dirac potentials
explicitly. This is an essential difference between the
relativistic and
non-relativistic approaches, and the presence of these potentials
can lead
to large differences in the observables obtained via relativistic
and non-relativistic approaches.
In order to obtain the usual non-relativistic expression, the
Dirac potentials must be removed from the nuclear current operators.
When this is done, the non-relativistic equivalent of the S-matrix
of Eq.  (\ref{f13})
yields the usual first order non-relativistic
transition amplitude used by many authors \cite{BGP,ML}.
When terms to second order are included in the non-relativistic
S-matrix, and in the
limit of no nuclear potentials, there are some differences between
our expression
and the usual non-relativistic second order S-matrix, which is
obtained via a
Foldy-Wouthuysen transformation of the interaction between
electrons and free nucleons \cite{BGP}.
Fearing, Poulis and Scherer \cite{FPS94} have compared
Foldy-Wouthuysen and
Pauli reductions of a Dirac hamiltonian containing a generic
potential
with harmonic time dependence. They found that differences do occur
beyond first order in $1/M$.
Detailed calculations show that these differences between the
Pauli and Foldy-Wouthuysen reductions are small when the nuclear
potentials are ignored in the nuclear current operators.
This seems the only consistent way to compare the operators since
we use two different hamiltonians for the Pauli calculations.

We discuss below the effects that the
presence of the potentials in the non-relativistic current
operators have on calculated observables.

\section{Results of Non-Relativistic Calculations} \label{n-results}

We now discuss results of numerical calculations using the proper
non-rela\-tiv\-istic reduction presented above.
In Fig. 2. we show results for the reaction on an $^{16}O$ target
with the same kinematics as in Fig. 1.
Figure 2(a) shows the spectral function while 2(b) shows the
proton polarization.
The non-relativistic calculations show the same effects due to
the inclusion of the
nuclear potentials in the interaction operators that we saw in
the corresponding convergence calculations of Fig. 1.
The first and second order calculations without potentials in
the nuclear currents
(~dotted and dot-dashed curves respectively~) yield very similar
results.
This is generally true in the cases we have considered;
going from first to
second order in $1/M$ does little to move the results in the
direction of the relativistic calculations.
When potentials are included in the nuclear currents, a large
change is noticeable
in going from first order to second order calculations,
particularly at larger values of missing momentum.
Note that the
non-relativistic calculations for the spectral function converge
to a  lower value
than the simple expansion in powers of $\left( E + M \right)^{-1}$
(~i.e. below the relativistic calculation~).
This is because the normalization of the
Schr\"{o}dinger-equivalent bound state wave function to unity.
This results in the non-relativistic expansion converging at a
point which is not the
relativistic one, but a factor of the square of the inverse
normalization constant lower
than the fully relativistic result. This amounts to a reduction
of the spectral
function from the relativistic result by a factor typically in
the range 1.2 to 1.4.
Spin observables are not affected by changes in overall
normalization,
so the proton polarization calculations shown in Fig. 2(b) are
very similar to those
shown in Fig. 1(b), with slight differences coming from the
replacement
$E \rightarrow M$ in the non-relativistic nuclear current
operators.

Figure 3 emphasizes the behavior of  the spectral function for
the high missing
momentum  region of Fig. 2(a), with the missing momentum in the
range 150 MeV/c to 300 MeV/c.
In this region the first and second order calculations without
potentials lie above the
relativistic calculations, while the inclusion of potentials
in first and second order
moves the results to lie below the relativistic results.
Note that the the relativistic calculations were fitted to the
data in the low missing
momentum region, but still do rather well for high missing momenta.
Similar results are obtained for $^{40}Ca$ and $^{90}Zr$ targets,
whether the residual nucleus is left in the ground or excited state.
When the potentials are not included in the nuclear currents the
results diverge
from the relativistic calculations as the magnitude of the missing
momentum is
increased. On the other hand, calculations which include potentials
in the
nuclear current operators remain close to the relativistic results
over a wide
range of missing momenta. Note that we are only including terms to
second order in the inverse mass.

Figure 4 shows non-relativistic calculations of the spectral
function and proton polarization for the same reaction discussed
in the previous figures, however in this case the energy of the
incident electron is 2000 MeV, and the
kinetic energy of the detected proton is fixed at 400 MeV.
The larger energies allow for a much larger range of missing
momenta than  considered previously.
It is important to note that the first and second order
calculations of the spectral
function, without potentials included in the nuclear current
operators,
differ from the relativistic calculations by up to an order of
magnitude for large
missing momenta, while inclusion of the nuclear potentials results
in convergence
to the fully relativistic results in the high missing momentum
region.
In addition we see that for low missing momenta the
convergence point  is lower than the relativistic (~see insert~).
The $(~\gamma, p~)$ reaction shows behavior consistent with these
observations
for $(~e, e'p~)$ at high missing momentum \cite{HH}. The momentum
transfer in the
$(~\gamma, p~)$ reaction is generally in the range 400 MeV/c to
600 MeV/c
so these two reactions can both probe this part of the single
particle bound state wave function.

Proton polarization is shown in Fig. 4(b). In the region of large
missing
momentum there are large differences between the polarization
calculated
with and without nuclear potentials in the current operator.
The polarization calculated with first and second order currents
containing
nuclear potentials yields results close in magnitude and shape to
the results of the fully relativistic calculations.
Note in particular, that in the region of the minimum and maximum
in the relativistic
calculations close to
$p_{m} = -400 MeV/c$ and $p_{m} = 400 MeV/c$ respectively,
the calculations without potentials included do not reproduce
the shape of the relativistic calculations at all.
The potentials must be included in the nuclear current operators
in order to get close to the relativistic results.
In particular a measurement of the proton polarization near
$p_{m} = -400 MeV/c$ provides a clear opportunity to
differentiate between relativistic and non-relativistic models.

Calculation of the asymmetry parameter of Eq. (\ref{f15}),
for small values of angular momentum
$L$ of the bound nucleon (~$L \leq 2$~),  yields similar results
for all the calculations whether fully relativistic; or
non-relativistic first or second order,
with or without potentials in the nuclear current operator.
When the angular momentum of the bound nucleon is increased,
the differences
between these calculations of the asymmetry become larger,
as is evident in  Fig. 5.
The asymmetry is calculated  for a $^{90}Zr$ target, with the
residual state in $^{89}Y$ assumed to be a $1f_{\frac{5}{2}}$
proton hole.
The incident electron has an energy of 461 MeV, and the kinetic
energy of the
detected proton is fixed at 100 MeV.
In this case the differences are particularly apparent for
missing momenta in the neighborhood of 20 MeV/c.

\section{Conclusions}                   \label{concl}

In order to clarify the differences arising from relativistic
and non-relativistic
descriptions of quasifree electron scattering \cite{J,JU},
we have discussed an expansion of the S-matrix for the reaction
$(~e, e'p~)$ in
powers of $\left( E + M \right)^{-1}$ through the effective
Pauli scheme.
The resulting S-matrix depends on Schr\"{o}dinger-like wave
functions for the
bound and continuum nucleons, and nuclear current operators
which contain the
strong Dirac potentials at the different orders.
When the Dirac potentials are included in the nuclear current
operators, the series
essentially converges to the fully relativistic results at
second order for light- to
medium-mass nuclei we have considered.
This indicates the importance of the role played by the nuclear
potentials in the
modification of the currents. When the potentials are not included
in the nuclear
currents, the calculations can be far from the relativistic
results particularly for larger missing momenta.

These points were further studied in setting up a comparison
between relativistic and non-relativistic calculations.
A proper non-relativistic calculation is obtained through
several steps:
normalization of the bound Schr\"{o}dinger-like wave function
to unity,
proper normalization of the continuum Schr\"{o}dinger-like
wave function,
and in the nuclear current operators the energy of the nucleons
is set equal to the
nucleon mass ( i.e. take the limit  $E \rightarrow M$~).
An additional step of removing the Dirac potentials from the
resulting nuclear current operators yields the standard
non-relativistic amplitude.
This results in a consistent and fair comparison between the
relativistic and non-relativistic calculations.
The potentials used for the bound and continuum
protons yield both the relativistic and non-relativistic
wave functions, with normalizations handled appropriately.

The non-relativistic calculations we have shown for first and
second order
nuclear current operators without potentials give the same
results that a standard
non-relativistic calculation would give if provided with
the Schr\"{o}dinger
equivalent wave functions derived from the Dirac equation.
Inclusion of the nuclear potentials in the non-relativistic
nuclear current operator
results in a large change in the calculated observables.
In particular, calculations of the spectral function and final
proton polarization
using second order nuclear current operators which include the
Dirac potentials,
can reproduce the magnitude and shape of the fully relativistic
calculations.
This is true even at large missing momenta where the
non-relativistic calculations
without potentials in the nuclear current operators yield
very different results
than the fully relativistic calculations.
The polarization of the final proton is particularly sensitive
to differences in
the calculations, and measurements of this observable at large
missing momenta
could assist in the choice between the relativistic and
non-relativistic approaches.

We have also calculated the asymmetry defined in the text for
the different orders,
with and without potentials, and found that in
cases in which the angular momentum of the bound nucleon is less
than 2, there
are no noticeable differences between these calculations and the
full relativistic.
When the orbital angular momentum of the bound nucleon is greater
than 2
differences between  the resulting asymmetry in missing momentum
of these calculations will appear.
This observable thus will be useful in differentiating between
relativistic and
non-relativistic models only for nuclear states with large
orbital angular momentum.

Other groups have examined the sensitivity of the models to changes
in
the optical potentials and modifications of the wave functions
\cite{BGPC87,Ud93,JO94}, and have found sensitivities at the level
of 15\%.
However, the essential differences between the relativistic and
non-relativistic
approaches do not lie in modifications of the wave functions.
The essential
difference comes from the appearance of the nuclear potentials in
the nuclear current operators; a result of the reduction of the
relativistic amplitude.
We emphasize that {\em these nuclear medium effects,
characteristic of the present model, will not appear through a
non-relativistic impulse description of the process}.
They are, however, inherent in the relativistic description.

\section*{Acknowledgements}

We would like to thank J.M. Udias and L. Lapik\'{a}s for useful
correspondence regarding quasifree electron scattering.

\newpage

\section* {Figure Captions}

\noindent FIG. 1. Observables for the reaction
$^{16}O(~e, e'p~)^{15}N$ where
$^{15}N$ is in the $1p_{\frac{1}{2}}$ state.
The energy of the incident electron is 456 MeV, and the
kinetic energy of the
detected proton is fixed at 90 MeV with parallel kinematics.
Hartree bound state wave functions are used \cite{HS} and the
proton optical
potentials are from \cite{COPE}.
The data are from reference \cite{L}.
(a) spectral function and (b) proton polarization.
Curves labelled according to their order in
$\left( E + M \right)^{-1}$
and whether or not the Dirac potentials are included in the
nuclear current operators:
dotted curve -- first order in $\left( E + M \right)^{-1}$
without Dirac potentials;
dashed curve -- first order with potentials;
dot-dashed curve -- second order without potentials;
dot-dot-dashed curve -- second order with potentials;
solid curve -- fully relativistic calculation.

\vspace{2.5 mm}
\noindent FIG. 2. Observables for the reaction
$^{16}O(~e, e'p~)^{15}N$ where
$^{15}N$ is in a $1p_{\frac{1}{2}}$ proton hole state.
The kinematics are those of Fig. 1.
Curves labelled according to their order in $1/M$ and whether
or not the
Dirac potentials are included in the nuclear current operators:
dotted curve -- non-relativistic calculations, first order in
$\left( E + M \right)^{-1}$ without Dirac potentials;
dashed curve -- first order with potentials;
dot-dashed curve -- second order without potentials;
dot-dot-dashed curve -- second order with potentials;
solid curve -- fully relativistic calculation.
Potentials and data from the sources of Fig. 1.

\vspace{2.5 mm}
\noindent FIG. 3. Spectral function for $^{16}O(~e,~e'p~)^{15}N$
where $^{15}N$ is in the $1p_{\frac{1}{2}}$ state.
The kinematics are those of Fig. 1.
Curves labelled as in Fig. 2.
Potentials and data from the sources of Fig. 1.

\vspace{2.5 mm}
\noindent FIG. 4. Observables for $^{16}O(~e,~e'p~)^{15}N$ where
$^{15}N$ is in the $1p_{\frac{1}{2}}$ state.
The energy of the incident electron is 2000 MeV, and the kinetic
energy of the
detected proton is fixed at 400 MeV with parallel kinematics.
Curves labelled as in Fig. 2.
Potentials from the source of Fig. 1.

\vspace{2.5 mm}
\noindent FIG. 5. Asymmetry in missing momentum for
$^{90}Zr(~e,~e'p~)^{89}Y$
where $^{89}Y$ is in the $1f_{\frac{5}{2}}$ state.
The energy of the incident electron is 461 MeV, and the kinetic
energy of the
detected proton is fixed at 100 MeV with parallel kinematics.
Curves labelled as in Fig. 2.
Potentials from the source of Fig. 1.

\begin{figure}
\begin{picture}(1100,400)(0,0)
\includegraphics{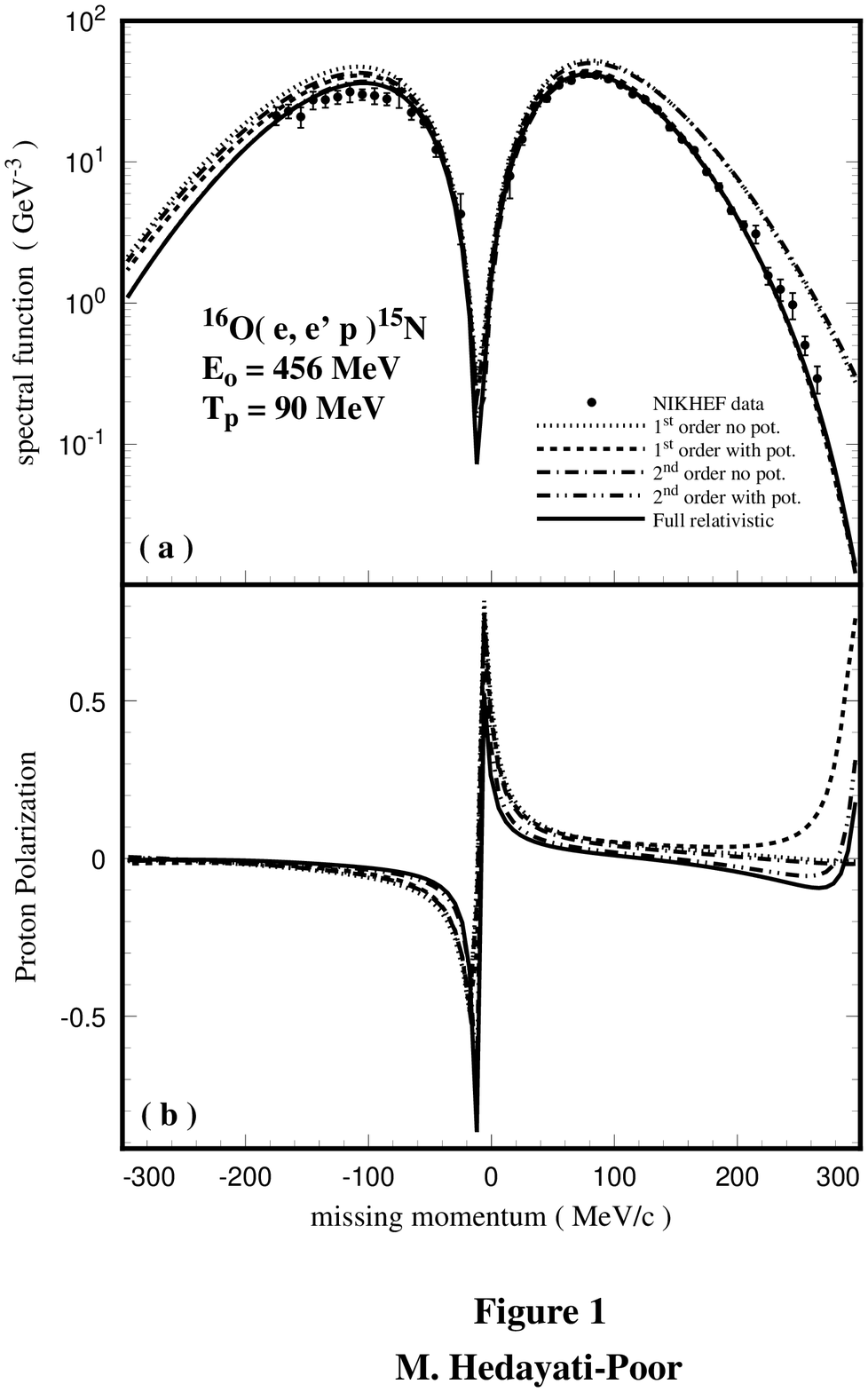}
\end{picture}
\end{figure}

\begin{figure}
\begin{picture}(1100,400)(0,0)
\includegraphics{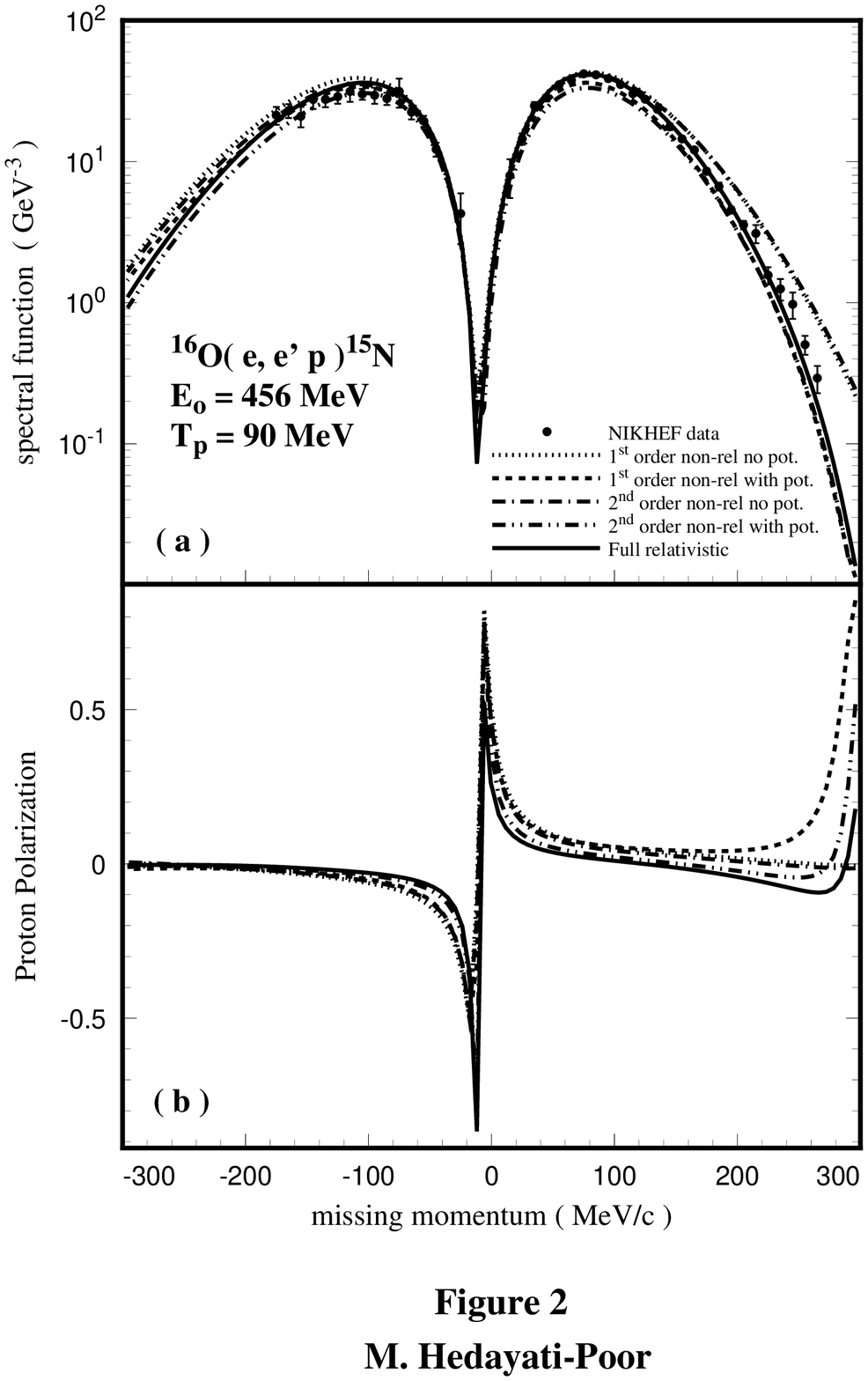}
\end{picture}
\end{figure}

\begin{figure}
\begin{picture}(1100,400)(0,0)
\includegraphics{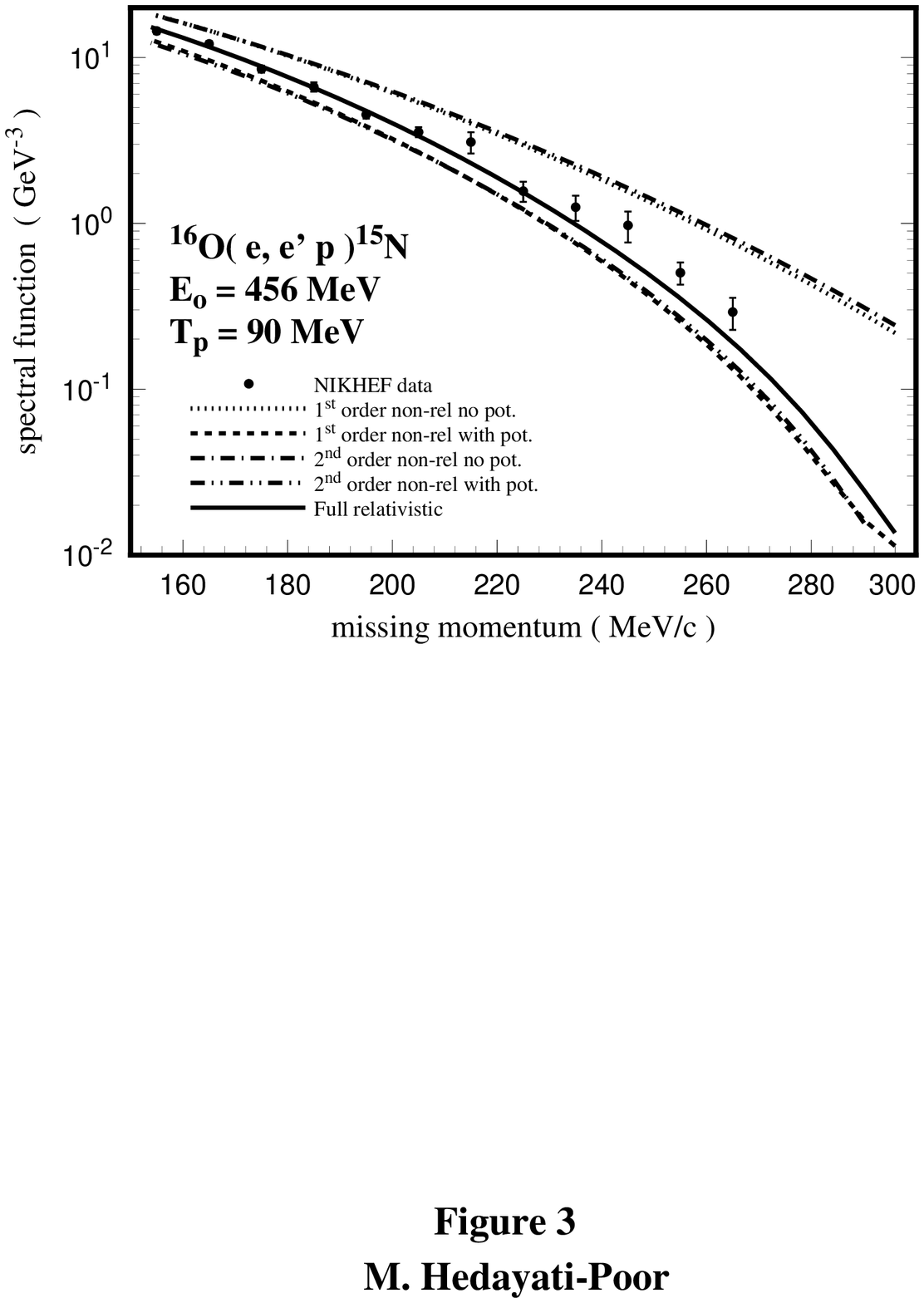}
\end{picture}
\end{figure}

\begin{figure}
\begin{picture}(1100,400)(0,0)
\includegraphics{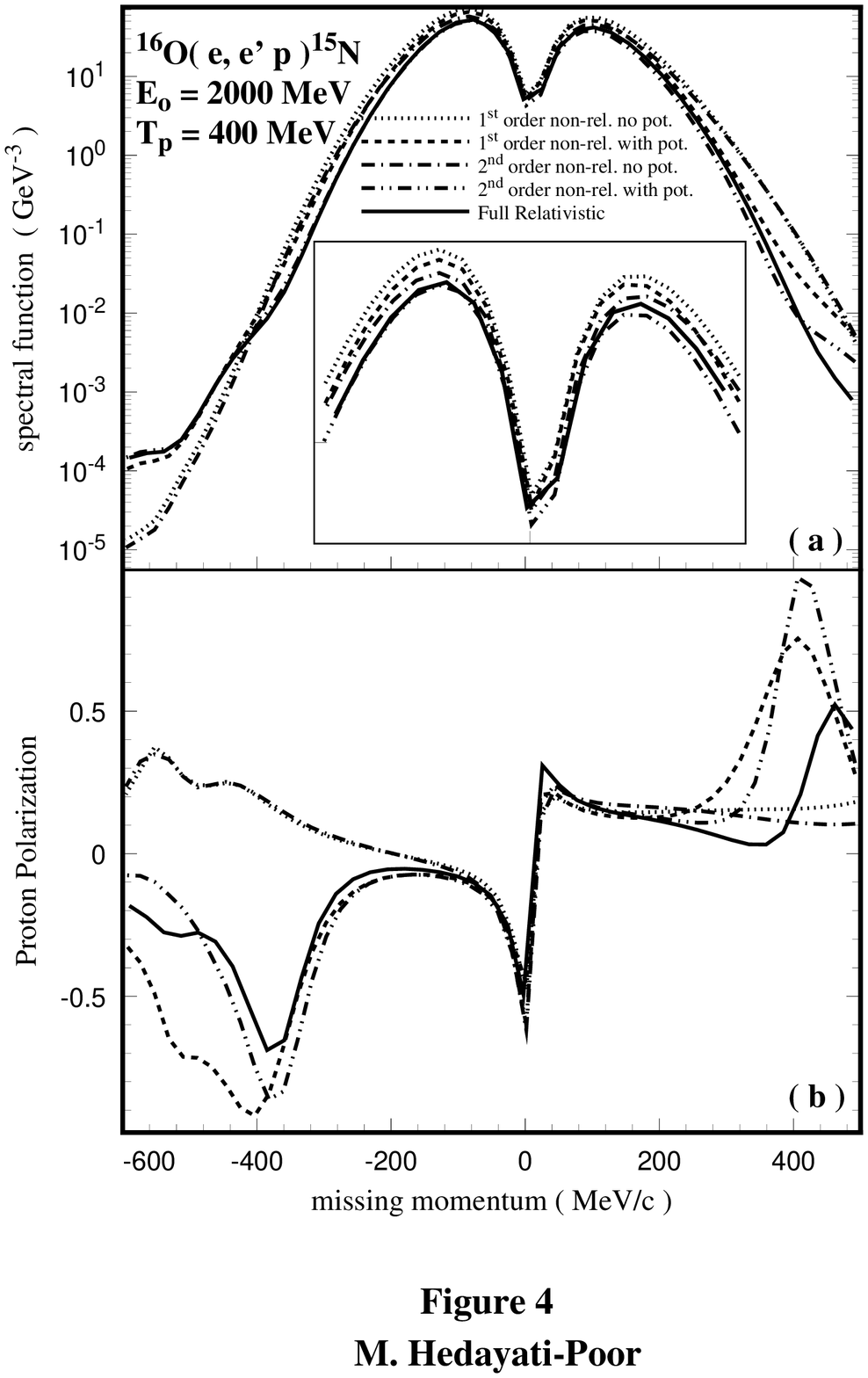}
\end{picture}
\end{figure}

\begin{figure}
\begin{picture}(1100,400)(0,0)
\includegraphics{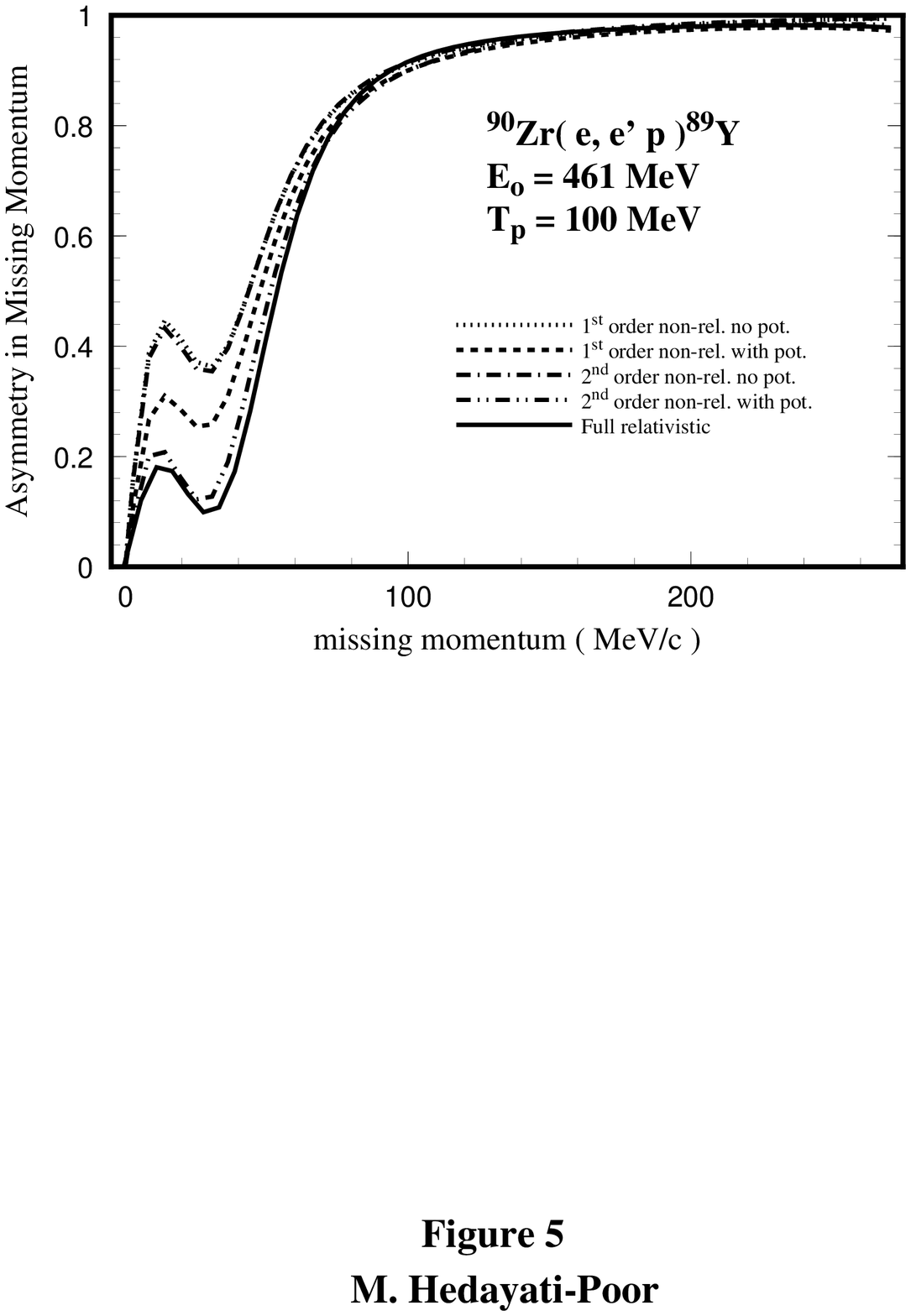}
\end{picture}
\end{figure}

\end{document}